\documentclass{aa}
\usepackage{graphics,graphicx,epsfig,psfig}
\usepackage{times}
\input{epsf.sty}

\begin{document}


   \title{On the intensity contrast of solar photospheric faculae and network
elements}

   \author{A. Ortiz\inst{1}
          \and
          S.K. Solanki\inst{2}
	  \and
	  V. Domingo\inst{1,3}
	  \and
	  M. Fligge\inst{4}
	  \and
	  B. Sanahuja\inst{1,3}
          }

   \offprints{A. Ortiz}
   \mail{aortiz@am.ub.es}

   \institute{Departament d'Astronomia i Meteorologia, Universitat de
              Barcelona, Mart\'{\i} i Franqu\`{e}s 1, E-08028 Barcelona, Spain
	      \and
	      Max-Planck-Institut f\"ur Aeronomie, Max-Planck-Str. 2, D-37191 			      Katlenburg-Lindau, Germany
	      \and
	      Institut d'Estudis Espacials de Catalunya, Gran Capit\`{a} 2-4, 		      	      E-08034 Barcelona, Spain	 
	      \and
	      Institute of Astronomy, ETH-Zentrum, Scheuchzerstr. 7, CH-8092 			      Z\"urich, Switzerland
             }

   \date{Received / Accepted}
   
   \abstract{    %
   Sunspots, faculae and the magnetic network contribute to
solar irradiance variations. The contribution due to faculae and the network is
of basic importance, but suffers from considerable uncertainty. We determine
the contrasts of active region faculae and the network, both as a function of
heliocentric angle and magnetogram signal. To achieve this, we analyze
near-simultaneous full disk images of photospheric continuum intensity and
line-of-sight magnetic field provided by the Michelson Doppler Interferometer
(MDI) on board the SOHO spacecraft. Starting from the surface distribution of
the solar magnetic field we first construct a mask, which is then used to
determine the brightness of magnetic features, and the relatively field-free
part of the photosphere separately. By sorting the magnetogram signal into
different bins we are able to distinguish between the contrasts of different
concentrations of magnetic field. We find that the center-to-limb variation
(CLV) of the contrast changes strongly with magnetogram signal. Thus, the
contrasts of active region faculae (large magnetogram signal) and the network
(small signal) exhibit a very different CLV, showing that the populations of
magnetic flux tubes that underly the two kinds of features are different. The
results are compatible with, on average, larger flux tubes in faculae than in
the network. This implies that these elements need to be treated separately
when reconstructing variations of the total solar irradiance with high
precision. We have obtained an analytical expression for the contrast of
photospheric magnetic features as a function of both position on the disk and
spatially averaged magnetic field strength, by performing a 2-dimensional fit
to the observations. We also provide a linear relationship between magnetogram
signal and the $\mu=\cos(\theta)$, where $\theta$ is the heliocentric angle, at
which the contrast is maximal. Finally, we show that the maximum contrast per
unit magnetic flux decreases rapidly with increasing magnetogram signal,
supporting earlier evidence that the intrinsic contrast of magnetic flux tubes
in the network is higher. %
}

\titlerunning{On the contrast of solar photospheric faculae and network elements}

   \maketitle


\section{Introduction}
\label{intro}

Radiometers on board satellites launched during the last three decades
(NIMBUS-7, SMM, UARS, EURECA, SOHO) have revealed that the total solar
irradiance, also referred to as the solar constant, changes on a variety of
time-scales. Solar irradiance variations on scales of days up to the solar
activity cycle length are closely related to the evolution of the solar surface
magnetic field, because the emergence and evolution of active regions (AR) on
the solar surface is reflected in the irradiance records (Lean et al.
\cite{lean98}; Fligge \& Solanki \cite{fligge2000b}). Sunspots and active
region faculae are considered to be the dominant contributors to solar
irradiance changes on time-scales of days to weeks. Space-based irradiance
records have also established a variation of about 0.1\% of the irradiance in
phase with the 11 year solar activity cycle, giving as a result a brighter Sun
around activity maximum (Chapman \cite{chap87}; Willson \& Hudson
\cite{willhud88}). The origin of the long-term increase of the irradiance
between activity minimum and maximum is still widely debated. It is expected
that small-scale magnetic elements that compose the enhanced and quiet network
contribute substantially to the observed irradiance increase during activity
maximum (Foukal \& Lean \cite{foulean88}; Solanki \& Fligge \cite{solfli01};
Fligge \& Solanki \cite{fligge2000a}). Nevertheless, other mechanisms of
non-magnetic origin have also been proposed, based, for example, on temporal
changes in the latitude-dependent surface temperature of the Sun (Kuhn et al.
\cite{kld88}). Other authors have tried to explain these variations by
modelling structural changes in the convection zone during the solar cycle
(Balmforth et al. \cite{gough96})

The photospheric magnetic field is concentrated in discrete elements whose
diameters range from less than a hundred to several tens of thousands of
kilometers. The brightness signature of these magnetic features is a strong
function of their heliocentric angle and their size; sunspots are dark while
small flux tubes are ge\-ne\-ra\-lly bright; faculae appear brighter near the
limb (Solanki \cite{solanki93}, \cite{solanki01}). However, our knowledge of
the brightness of small scale magnetic features, groups of which form faculae
and the network, is incomplete (e.g. Solanki \cite{solanki94}). 

Faculae, bright structures seen in the photosphere cospatially with
chromospheric plages, are associated with magnetic fields. At high resolution,
they consist of many unresolved small continuum bright points, with diameters
of about 100 km, called facular points (Muller \cite{muller83}; Berger et al.
\cite{berger95}). The observed zoo of small magnetic features is unifyingly
described by the concept of the small flux tube. To predict their physical
properties, different models for small flux tubes have been constructed (e.g.,
Spruit \cite{spruit76}; Deinzer et al. \cite{deinz84a}, \cite{deinz84b};
Kn\"olker et al. \cite{ketal88}; Kn\"olker \& Sch\"ussler \cite{ks88};
Grossmann-Doerth et al. \cite{grossdo89}; Steiner et al. \cite{steiner96}).
According to this model faculae are conglomerates of evacuated flux tubes with
hot walls and a hot or cool floor (corresponding to an optical depth of
$\tau=1$) depending on the evacuation and diameter of the flux tube. The model
predicts a certain CLV of the contrast for a particular diameter of the
underlying flux tubes.

This model assumes that inside each small flux tube the magnetic field is of
the order of a kilogauss, but practically zero outside. Due to the magnetic
pressure the flux-tube interior is evacuated, so that $\tau=1$ is reached along
its walls (which are bright due to radiation leaking in from the surroundings,
i.e., there is a horizontal flux of energy into the tube). Flux tubes can be
dark at disk center if suppression of convective energy transport within the
tube is included (e.g., Spruit \cite{spruit76}; Deinzer et al. \cite{deinz84a},
\cite{deinz84b}; Kn\"olker et al. \cite{ketal88}), resulting in a cooling of
the deeper layers. When the tube is sufficiently broad, the horizontal optical
depth between the wall and the tube center is large and most of the radiation
cannot reach the center. In this case, the tube floor remains dark at its
center. But if the tube is sufficiently slender, the horizontal flux of energy
can reach the center of the tube; then, the interior of the tube is heated, the
vertical energy flux increases, and the tube turns bright, even at disk center.
In this scenario, the transition between smaller bright points and larger dark
micropores (e.g., Topka et al. \cite{topka97}) occurs at a diameter of about
300 km (e.g., Grossmann-Doerth et al. \cite{grossdo94}), and therefore
micropores would fill the gap between small bright points and larger dark
pores. Micropores are predominantly found in active regions, while bright
points are the main constituents of the network. When observed near the limb,
the heated walls of the tube become visible, and therefore the contrast
increases.

Observations also provide evidence that the contrast, as well as the underlying
thermal structure, depends on the size of the flux tubes (e.g., Keller
\cite{keller92}), but more commonly on the strength of the magnetogram signal
(Frazier \cite{fra71}; Spruit \& Zwaan \cite{spruitz81}; Solanki \& Stenflo
\cite{solstenf84}; Solanki \cite{solanki86}; Zayer et al. \cite{zayer90};
Solanki \& Brigljevi\'{c} \cite{solbri92}; Topka et al. \cite{topka92},
\cite{topka97}; Lawrence et al. \cite{lawtopjo93}; Grossmann-Doerth et al.
\cite{grossdo94}), and hence a test of flux tube models is possible with such
data. However, since most flux tubes are not resolved, it is necessary to have
a well-defined and constant spatial resolution of the observations with which
to compare the models. With ground-based data, the basis of practically all
facular contrast CLV measurements to date, this criterion is hard to meet.

Differences in spatial resolution, caused by variable seeing, may indeed partly
explain the variety of measured contrast CLVs (e.g., Libbrecht \& Kuhn
\cite{lk84}; Unruh et al. \cite{unruh99}). Other possible factors are
differences in wavelength, spectral resolution and the magnetic filling factor
of the observed features (Solanki \cite{solanki94}). The problem posed by
variable seeing can be circumvented by employing data recorded in space, while
the magnetic filling factor can be estimated with the help of cospatial and
cotemporal magnetograms. Only relatively few contrast investigations including
the magnetogram signal can be found in the literature (e.g. Frazier
\cite{fra71}; Foukal \& Fowler \cite{ff84}; Topka et al. \cite{topka92},
\cite{topka97}; Lawrence et al. \cite{lawtopjo93}).

Here we add to this list using data from the MDI instrument on board SOHO
(Domingo et al. \cite{dom95}); their main advantages are:

\renewcommand{\labelitemi}{$-$}
\begin{itemize}
\item seeing effects due to the Earth's atmosphere are avoided,
\item measurements are made in the pure continuum (i.e. spectral resolution is not a problem),
\item the 20-minute averaged MDI magnetograms have a reasonably low noise level,
\item a large and homogeneous data set is available,
\item the characteristics of the instrument and the data sets are well known and stable,
\item magnetograms and intensity images are obtained regularly by the same
instrument with exactly the same spatial resolution, so that a one-to-one identification of brightness with magnetogram signals can be made.
\end{itemize}

The main disadvantages of the MDI data are:

\renewcommand{\labelitemi}{$-$}
\begin{itemize}
\item a relatively low spatial resolution with a pixel size of
$2\times2\arcsec$,
\item measurements are available at only a single wavelength.
\end{itemize}

The purpose of this paper is to present new high-quality measurements of the
contrast of the photospheric bright features as a function of both heliocentric
angle and magnetogram signal and to obtain an analytical function that predicts
their contrast given a position on the disk and a magnetic signal value. Such
measurements are expected to be of use not only to constrain models of flux
tubes, but also to improve the modelling of the solar irradiance (Lean et al.
\cite{lean98}). Uncertainties in the contrast of faculae and the network are
one of the major sources of error in the modelling of solar irradiance
variations. Employing MDI data to obtain the contrast as input for irradiance
modelling is of particular interest since MDI magnetograms have already been
succesfully used for such modelling (Fligge \& Solanki \cite{fligge2000a}).


In Sect.~\ref{dataproc} we present the data sets used and the analysis
procedures. In Sect.~\ref{results} we describe the results, which are discussed
in Sect.~\ref{discu}. Finally, our conclusions and a summary are given in
Sect.~\ref{conclu}.


\section{Data and analysis procedure}
\label{dataproc}
\subsection{Data sets}

The Solar Oscillations Investigation/Michelson Doppler Imager (SOI/MDI)
instrument is a state-of-the-art helioseismology experiment and magnetograph on
board the SOHO spacecraft, devoted to study the interior structure and dynamics
as well as the surface magnetic field of the Sun. This instrument gives an
image of the Sun on a $1024\times1024$ CCD camera, and can observe in two
spatial resolution modes, full disk and high-resolution of the central part of
the disk (HR). We are interested in the full disk measurements, which have a
field of view of $34\times34\mbox{\arcmin}$ and a pixel size of
$2\times2\mbox{\arcsec}$. Two tunable Michelson interferometers allow MDI to
record filtergrams centered at five wavelengths across the \ion{Ni}{i} 6768
\mbox{\AA} absorption line. From the filtergrams, MDI computes the following
six observables: Doppler velocity, continuum intensity, line depth,
longitudinal magnetic field, horizontal velocity and limb position. The SOI/MDI
instrument is described in detail by Scherrer et al. (\cite{sche95}).

The products of interest for our work are the full disk magnetograms and
continuum intensity images. Magnetograms only measure net magnetic flux per
resolution element, therefore the signal is not the true magnetic field
strength {\bf B}, inside a flux tube, but its longitudinal component, $\langle
|\bf B| \cos\gamma\rangle$, averaged over the pixel, where $\gamma$ is the
angle between the magnetic vector and the line of sight. For simplicity, we
hereafter refer to $\langle |\bf B| \cos\gamma\rangle$ as $B$. In a 2-component
model of the magnetic field, with magnetic flux tubes of field strength {\bf B}
covering a fraction $\alpha$ of the solar surface separated by a field-free
component covering $(1-\alpha)$ of the surface, we can write $\langle |\bf B|
\cos\gamma\rangle$ as $\alpha |\bf B| \cos\gamma$. Since the true field
strength $|\bf B|$ lies in a relatively narrow range of 1000-1500 \mbox{G} for
all magnetic features except intranetwork elements (Solanki et al.
\cite{solfins99}), and $\cos\gamma\approx\cos\theta\approx\mu$ is a reasonable
approximation ($\theta$ is the heliocentric angle), the strength of the
magnetogram signal mainly provides information on the magnetic filling factor
$\alpha$. 

MDI magnetograms are usually obtained every 96 minutes, with the exception of
periodic campaigns in which 1-minute cadence measurements are available. The
1-$\sigma$ noise level for a one-minute longitudinal magnetogram is 20
\mbox{G}. Full disk continuum intensity images are taken each minute with a
noise level of 0.3\%.

The analyzed data set consists of nearly simultaneous magnetograms and
continuum intensity images recorded on 10 days in the period February to
October, 1999, as shown in Table~\ref{table1}. The time of the observations is
given for the averaged magnetograms (see Sect.~\ref{reduction}). It corresponds
to the middle of the 20 minute integration time. These days were chosen because
they belong to a high activity period so that everything from quiet network to
intense active-region plage was present on the solar surface. The sample
contains active regions spread over almost all $\mu=\cos\theta$ values. They
are also generally well separated in time in order to avoid duplication.

\begin{table}
\caption{Selected days and times (hours/min/sec) during 1999 at which the
averaged magnetograms analyzed here were recorded (see text for details).}
\label{table1}
{\centering \begin{tabular}{ll}
\hline 
1999 observation dates & Time (UT)\\
\hline 
\hline
February 13 & 00:10:02 \\
February 20 & 04:10:02 \\
May 14 & 00:10:03 \\
May 28 & 06:10:03 \\
June 25 & 01:10:02 \\
July 2 & 03:10:02 \\
July 10 & 01:10:02 \\
August 7 & 00:10:02 \\
October 12 & 09:10:03 \\
October 15 & 06:10:03 \\
\hline 
\end{tabular}\par}
\end{table}

\subsection{Reduction method and analysis}
\label{reduction}

\begin{figure}
\vspace{0.7cm}
\hspace{-3.cm}
\resizebox{15cm}{19cm}{\includegraphics{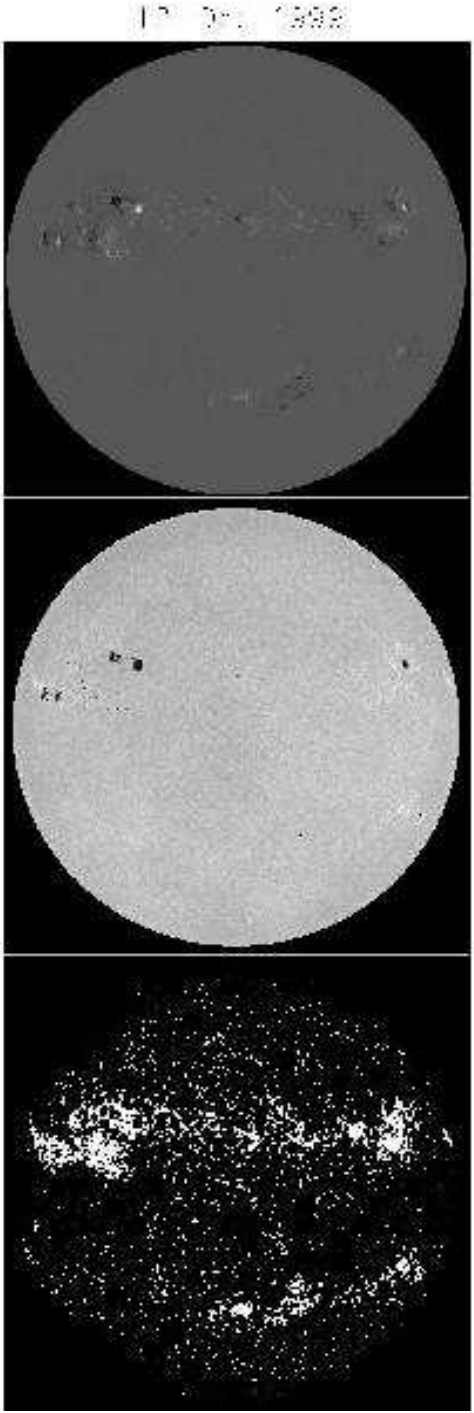}}
\caption{Example of a 20 minute averaged MDI magnetogram (top panel), the
corresponding intensity image after removal of limb-darkening (middle panel)
and the resulting contrast mask (lower panel) for October 12, 1999.}
\label{fig1}
\end{figure}

We employ averages over 20 single magnetograms, taken at a cadence of 1 per
minute, in order to reduce the noise level sufficiently to reliably identify
the quiet network. The individual magnetograms were rotated to compensate for
the time difference before averaging. Intensities are standard 1-minute images.
Care has been taken to use intensity images obtained as close in time to the
magnetograms as possible. In all cases but one, the two types of images were
recorded within 30 minutes of each other, with 37 minutes being the highest
difference. The intensity images have been rotated to co-align them with the
corresponding average magnetogram. Intensity images have also been corrected
for limb-darkening effects using a fifth order polynomial in $\mu$ following
Neckel \& Labs (\cite{nl94}). Our final data sets are pairs of co-aligned
averaged magnetograms and photospheric continuum intensity images for each of
the 10 selected days. Both types of images can be compared pixel by pixel. An
example magnetogram and the corresponding intensity image recorded on October
12, 1999 are shown in Fig.~\ref{fig1} (top and middle panels).

\begin{figure}
\vspace{-0.8cm}
\resizebox{\hsize}{!}{\includegraphics{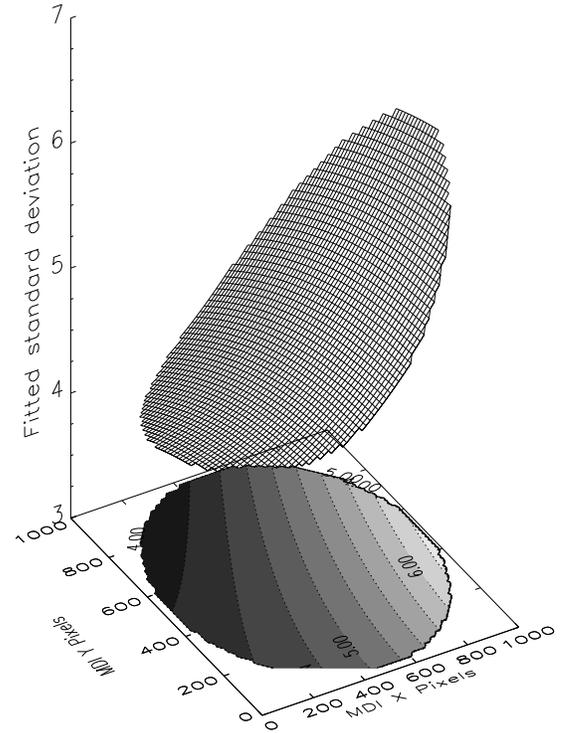}}
\caption{Standard deviation (in G) of the 20-minute averaged MDI magnetograms.
An increase of the noise in the direction of the SW limb is evident. The
shadowed contours indicate some of the values. Both the surface and the
contours represent the standard deviation.}
\label{fig2}
\end{figure}

We have determined the noise level of the MDI magnetograms and continuum images
as a function of position over the CCD array. The standard deviation for the
magnetic signal has been calculated using a running $100\times100$ pixel box
over the solar disk, with the exception of the limbs, which were avoided by
masking out an outer ring of 75 pixels width. This process was applied to
several 1996 low activity magnetograms, in order to avoid artifacts introduced
by the presence of active regions. After that, their median was determined to
eliminate the possible remaining activity. A second order surface was then
fitted to the result and extrapolated to cover the whole solar disk. The
resulting noise level, $\sigma_{\mathrm{mag}}$, shows an increase towards the
SW limb that probably includes some velocity signal leakage. In Fig.~\ref{fig2}
we show the calculated standard deviation for the 20-minute averaged
magnetograms. Note that when applying this noise surface to our data we have
assumed that the MDI noise level has remained unchanged between 1996 and 1999. 

A similar procedure has been used to determine the mean and standard deviation
of the quiet Sun continuum intensity for each selected day, $\langle
I_{\mathrm{qs}}\rangle$ and $\sigma_{\mathrm{Iqs}}$ respectively, where the
subscript $\mathrm{qs}$ denotes ``quiet Sun". Every pixel in the running mean
box with an absolute magnetic signal value below 0.5 times
$\sigma_{\mathrm{mag}}$ (i.e. pixels with corresponding magnetogram signal
between approximately $-2.5$ and 2.5 \mbox{G}) has been considered as a quiet
Sun pixel. 

The surface distribution of solar magnetic features that produce a bright
contribution to irradiance variations, is identified by setting two thresholds
to every magnetogram-intensity image pair. The first threshold looks for
magnetic activity of any kind, and is set to $\pm3\sigma_{\mathrm{mag}}$, which
corresponds, on average, to 15 \mbox{G}. As we are only interested in bright
magnetic features, the second threshold masks out sunspots and pores by setting
all pixels with a continuum intensity $3\sigma_{\mathrm{Iqs}}$ below the
average to a null value. To reduce false detections, even at the risk of
missing active pixels, we reject all isolated pixels above the given thresholds
assuming that they are noise. 3 10$^4$ out of 10$^7$ analyzed data points are
rejected in this way. After this step, we find that 6\% of the pixels satisfy
both criteria. Using both thresholds we construct a mask of the contrast of
bright features for each day. The result of applying the mask derived from the
magnetogram (top panel) and intensity image (middle panel) shown in
Fig.~\ref{fig1}, is displayed in the bottom panel of that figure. Note that
only features that lie above the given thresholds in the magnetogram and the
intensity image are indicated by white pixels. Sunspots near the NE limb, for
example, do not appear in the mask, but faculae surrounding those sunspots are
well identified. Smaller features belonging to the magnetic network are also
pinpointed outside of the active regions, although weaker elements of the
network may well be missed. For each pixel with coordinates $(x,y)$, the
contrast $C_{\mathrm{fac}}$ is defined as: 

\begin{equation} 
C_{\mathrm{fac}}(x,y) = \frac{I(x,y) -
\langle I_{\mathrm{qs}}\rangle(x,y)}{\langle I_{\mathrm{qs}}\rangle(x,y)}.
\end{equation}

Contrast, magnetic field strength averaged over the pixel and position,
represented by the heliocentric angle $\mu=\cos(\theta)$, are calculated for
each selected pixel. Finally, for each of these parameters the pixels above the
thresholds for each of the 10 selected days are put together into vectors of
about 6 10$^5$ elements, which should provide adequate statistics for a
detailed study of the facular and network contrast.

The method used in this work resembles that employed by Topka et al.
(\cite{topka92}, \cite{topka97}), although our magnetic threshold is much lower
due to the less noisy magnetograms used. The angular resolution, however, is
also considerably lower, but it is constant.


\section{Results}
\label{results}

We have analyzed the AR faculae and network contrast dependence on both $\mu$
and the measured magnetic signal, $B$. It is important to recall that the
observed magnetic signal drops to zero at the limb, even if strong magnetic
field regions are present. This is a straightforward consequence of the fact
that magnetograms are only sensitive to the line-of-sight component of the
magnetic field and that the magnetic field is mainly vertical. To compensate
this effect to first order we have worked with $B$/$\mu$ (i.e. $\langle |\bf B|
\cos\gamma\rangle$$/\cos\theta$) instead of $B$. Second order effects due to
radiative-transfer effects or finite thickness of flux tubes remain.

\begin{figure*}
\hfill
\resizebox{17cm}{19cm}{\includegraphics{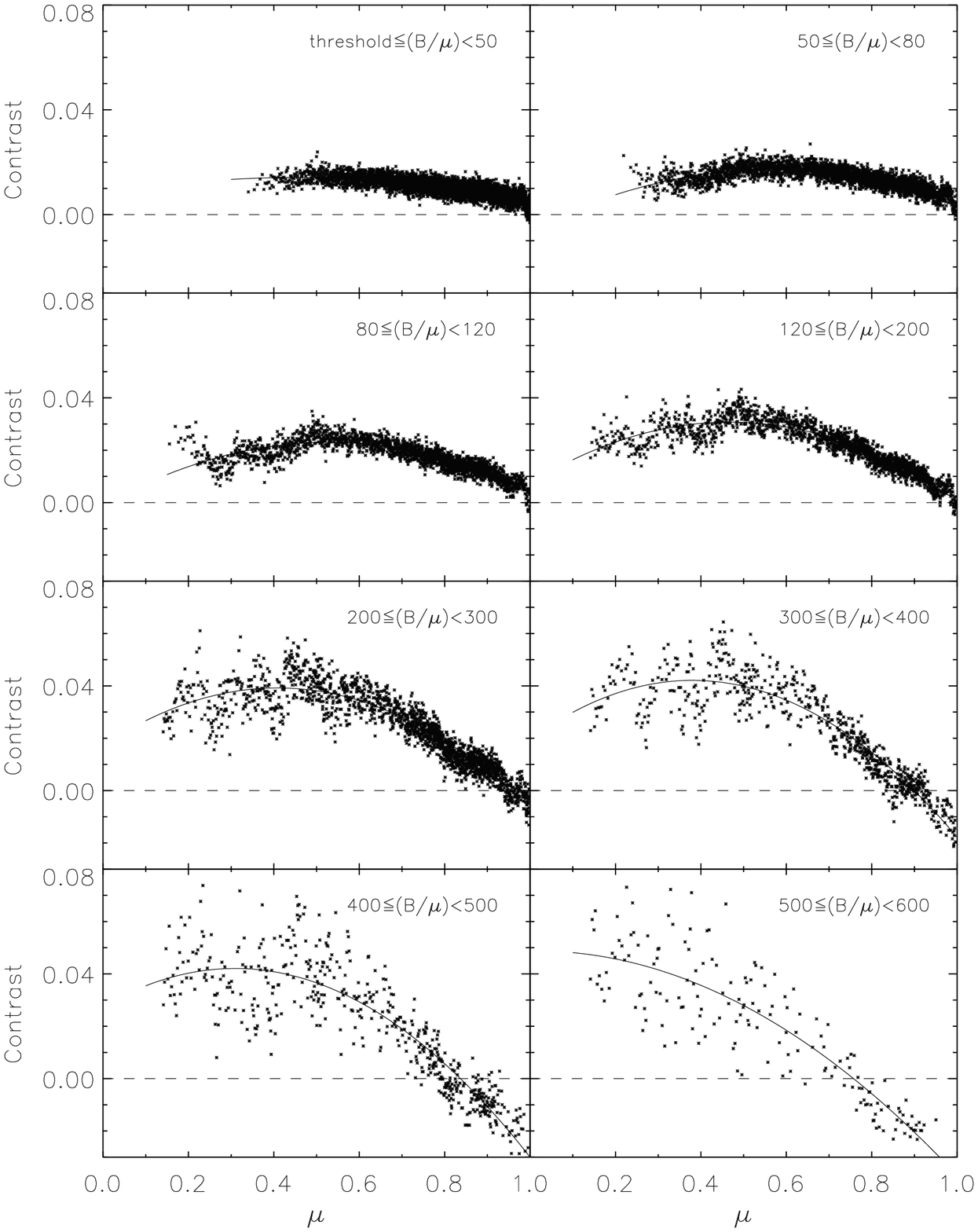}}
\vspace{1.5cm}
\caption{Facular and network contrast as a function of $\mu$ for eight intervals of the strength of the magnetic field, from network values (top left panel) to strong faculae (lower right). A dashed line at $C_{\mathrm{fac}}=0$ has been plotted. The solid curves represent a second order polynomial least-squares fit to the points. Every dot represents 40 data points. $\mu=1$ is the disk center; $\mu=0$ is the limb.}
\label{fig3}
\end{figure*}

\begin{figure*}
\hfill
\resizebox{17cm}{19cm}{\includegraphics{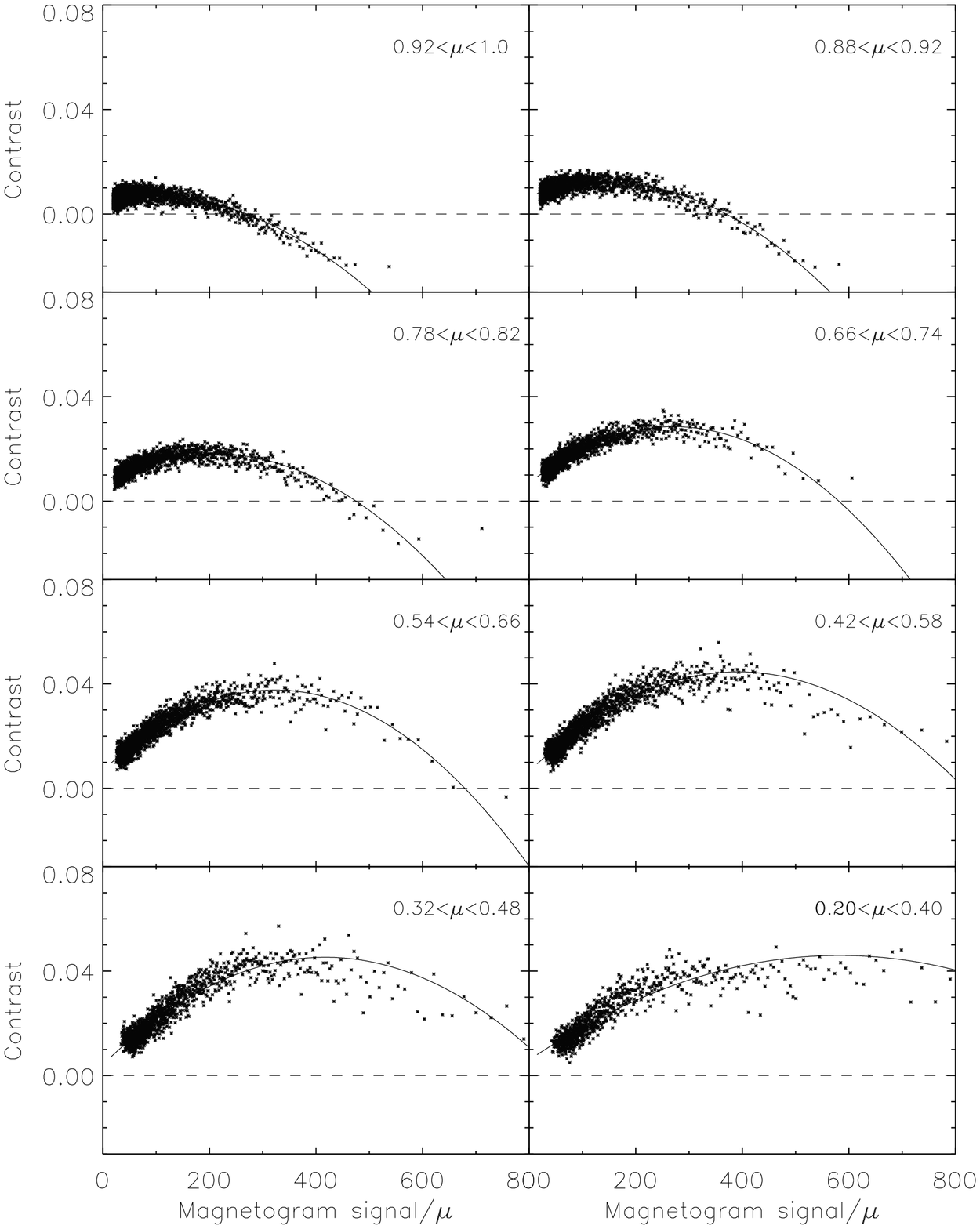}}
\vspace{1.5cm}
\caption{Dependence of the contrast on the absolute value of the magnetogram
signal, corrected for foreshortening effects. The solar disk has been divided
into eight bins, from center to limb. Note that some $\mu$-bins overlap (see
text for details). As in Fig.~\ref{fig3}, a dashed line at $C_{\mathrm{fac}}=0$
has been plotted and solid curves represent a second order polynomial
regression. Every dot represents 40 data points.}
\label{fig4}
\end{figure*}

We have binned the $B$/$\mu$ values into eight intervals that range from the
threshold level of, on average, 15 \mbox{G} to about 600 \mbox{G}. The
intervals have been chosen so that they roughly contain the same number of
pixels each, with the exception of the highest $B$/$\mu$ bins which have fewer
pixels. By sorting the magnetic field strength into different bins we can
distinguish between the CLV of the magnetic features present in regions with
different filling factor $\alpha$. Fig.~\ref{fig3} displays the contrast,
$C_{\mathrm{fac}}$, as a function of $\mu$, individually for every $B$/$\mu$
interval. A second order polynomial has been fitted (employing least squares)
to guide the eye and a dashed line indicating $C_{\mathrm{fac}}=0$ has been
included for clarity. To avoid overcrowding we have binned data points in sets
of 40 before plotting. 

Fig.~\ref{fig3} reveals a clear evolution of the behaviour of the contrast from
one $B$/$\mu$ interval to another. Network features (top left panel) show a low
and almost constant contrast, as compared with the very pronounced CLV of the
contrast for active region faculae (bottom panels). Intermediate cases show a
progressive increase of the contrast towards the limb as well as an
increasingly pronounced CLV. When $B/\mu < 200$ \mbox{G}, the contrast peaks at
$\mu\sim0.5$, while for magnetic signals $B$/$\mu\geq200$ \mbox{G} this maximum
shifts to lower values of $\mu$ (see Fig.~\ref{fig8} and its discussion in the
text for a more quantitative analysis). Note also that for $B$/$\mu\geq200$
\mbox{G} the contrast is negative around disk center, while it is positive in
the network (i.e. for smaller $B/\mu$). We will return to this point in
Sect.~\ref{discu}. The large fluctuations of the contrast near the limb for
intermediate and high magnetic signals are due to the distribution of active
regions on the ten selected days. 

Fig.~\ref{fig4} shows the contrast as a function of $B$/$\mu$, for different
positions on the solar disk. The solar disk has been divided into eight bins of
$\mu$, centred on $\mu$ values ranging from 0.96 (disk center, top left) to 0.3
(limb, bottom right). Note that to keep the number of points in each bin
approximately equal, $\mu$-bins lying closer to the limb are wider than the
ones around $\mu=1$, showing therefore some overlap. Each point of the figure
is obtained by binning together 40 points of data with similar $B$/$\mu$, and
second degree polynomials have also been plotted, as in Fig.~\ref{fig3}.

This figure shows that, in general, the contrast initially increases with
$B$/$\mu$ before decreasing again. We expect it to continue decreasing for even
larger values of $B$/$\mu$ representing pores and sunspots. At large $\mu$ the
initial increase is small and the contrast basically decreases with $B$/$\mu$,
while at small $\mu$ it mainly increases. For $\mu=1$, points with large
$B$/$\mu$ show a negative contrast (as in the lower panels of Fig.~\ref{fig3}),
while points at the limb always have positive contrasts. The bins with
\mbox{$\mu>0.82$} do not display data for high magnetic signals, because their
intensity is below the intensity threshold.

Given the regular behaviour of the contrast as a function of $\mu$ and
magnetogram signal, $C_{\mathrm{fac}}(\mu, B/\mu)$, it seems appropiate to
search for an analytical expression for this dependence. We have performed a
multivariate analysis using a $(\mu, B/\mu)$ grid. The $\mu$ values have been
binned linearly, with $\Delta\mu=0.1$. $B$/$\mu$ bins have been chosen to be
equally spaced on a logarithmic scale, with $\Delta\log(B/\mu)=0.05$, in order
to compensate for the fact that magnetic signals are mostly concentrated
towards lower values (Figs.~\ref{fig3} and \ref{fig4}). The dimensions of the
grid are \mbox{$0.1\leq\mu\leq1$} and
\mbox{$17\,\mbox{G}\leq(B/\mu)\leq630\,\mbox{G}$}, resulting in a $9\times31$
bins grid. We do not consider points with $B$/$\mu > 600$ \mbox{G} to exclude
bright features that might belong to pores observed near the limb. We are aware
of the fact that 600 \mbox{G} is an arbitrary value for such a cutoff.

Each bin of the grid is defined by the averaged values of the contrast, $\mu$
and $B$/$\mu$, over all the data points of that bin. Although the curves in
Fig.~\ref{fig3} are only intended to guide the eye, they do reveal that second
order polynomials fit the contrast as a function of $\mu$ well. We have fitted
the bidimensional array of contrasts by a second order polynomial function of
$\mu$ and a cubic function of $B$/$\mu$ of the form:

\begin{equation} 
C_{\mathrm{fac}}(\mu, B/\mu) = \sum_{i,j} a_{j,i}\mu^{j}(\frac{B}{\mu})^{i},
\label{eq2}
\end{equation}

\noindent where $i$ runs from 1 to 3, $j$ runs from 0 to 2 and $a_{j,i}$ are
the coefficients of the fit. The result of the fit is a surface of second order
in position $\mu$ and third order in magnetic signal. The coefficients of the
multivariate fit $a_{j,i}$ are:

\begin{eqnarray} 
C_{\mathrm{fac}}(\mu, B/\mu) & = &
10^{-4}\left[0.48+9.12\mu-8.50\mu^{2}\right]\left(\frac{B}{\mu}\right)+ \label{eq3} \\
& & 10^{-6}\left[0.06-2.00\mu+1.23\mu^{2}\right]\left(\frac{B}{\mu}\right)^{2}+ \nonumber \\  & &10^{-10}\left[0.63+3.90\mu+2.82\mu^{2}\right]\left(\frac{B}{\mu}\right)^{3}. \nonumber
\end{eqnarray}

\begin{figure}
\resizebox{\hsize}{!}{\includegraphics{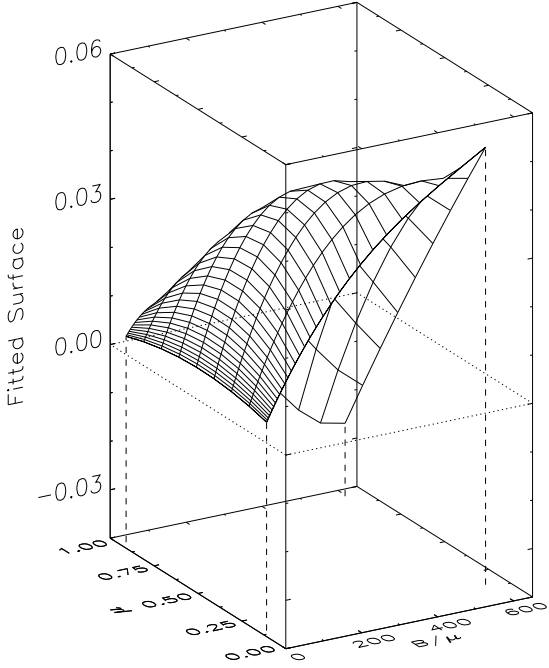}}
\caption{Polynomial surface of second order in $\mu$ and third order in
$B$/$\mu$ obtained from a multivariate fit performed to the grid of contrasts,
covering $\mu$ and $B$/$\mu$ values. Dashed vertical lines project the corners
of the plotted surface onto the $\mu$-$B$/$\mu$ plane and indicate the region
spanned by the fit.}
\label{fig5}
\end{figure}


The terms of Eq.~(\ref{eq2}) are grouped in Eq.~(\ref{eq3}) to make clearer the
quadratic dependence of $C_{\mathrm{fac}}(\mu, B/\mu)$ on $\mu$ and the cubic
dependence on $B/\mu$. When $B$/$\mu$ is small ($< 100$ G), the first order
term in $B$/$\mu$ provides the dominant contribution. When $B$/$\mu$ is large
($\geq 200$ G), first and second order terms in $B$/$\mu$ dominate the
contrast, modulated by the contribution of the cubic term which plays a role in
this range of magnetic signals. At disk center ($\mu=1)$, those terms result in
a dominant negative contribution (Fig.~\ref{fig3}). Note that the contrast is
constrained to go through zero when $B$/$\mu$=0, as expected for the quiet Sun.

The best-fit surface is shown in Fig.~\ref{fig5}. The grid corresponds to the
linear $\mu$-bins and the logarithmic $B$/$\mu$-bins taken for the fit. The
shape of this surface is quite congruent with that shown by the observed
contrast in Figs.~\ref{fig3} and \ref{fig4}. Note that the function given in
Eq.~(\ref{eq3}) is valid only for the wavelength and spatial resolution of the
MDI data, 6768 \mbox{\AA} and 2\arcsec\,, respectively. For other values of
these parameters we expect another dependence on $\mu$ and $B$/$\mu$. In
particular, the absolute value of the contrast is expected to change. 

\begin{figure}
\vspace{-0.75cm}
\leavevmode
\resizebox{\hsize}{9.75cm}{\includegraphics{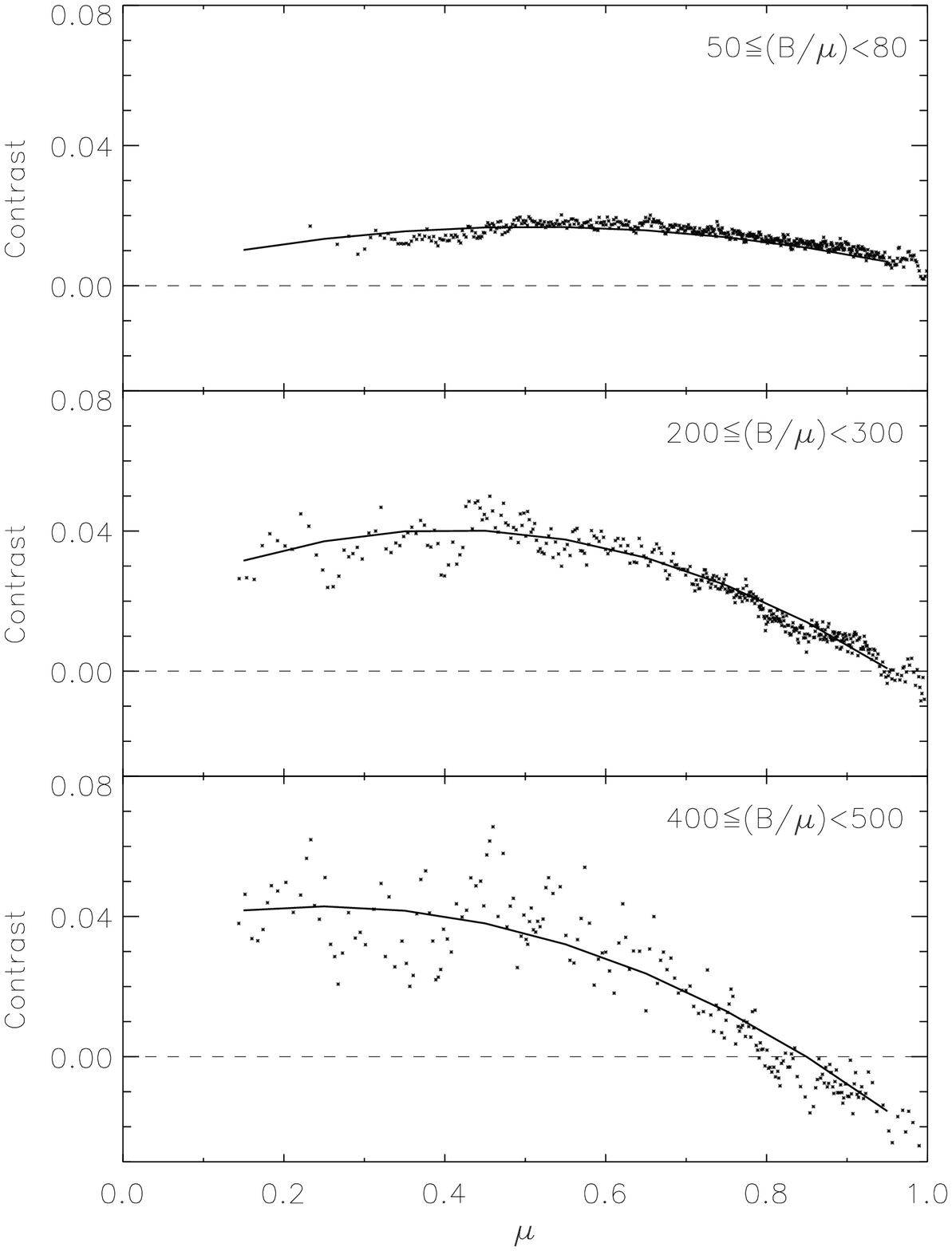}}
\vspace{0.2cm}
\caption{Comparison of cuts through the surface (solid curves) returned by the
multivariate analysis and the measured contrasts (dots) as a function of $\mu$,
for 3 sample bins of corrected magnetic signal. Every dot represents 250 (top),
100 (middle) and 25 (bottom) data points.}
\label{fig6}
\leavevmode
\resizebox{\hsize}{9.75cm}{\includegraphics{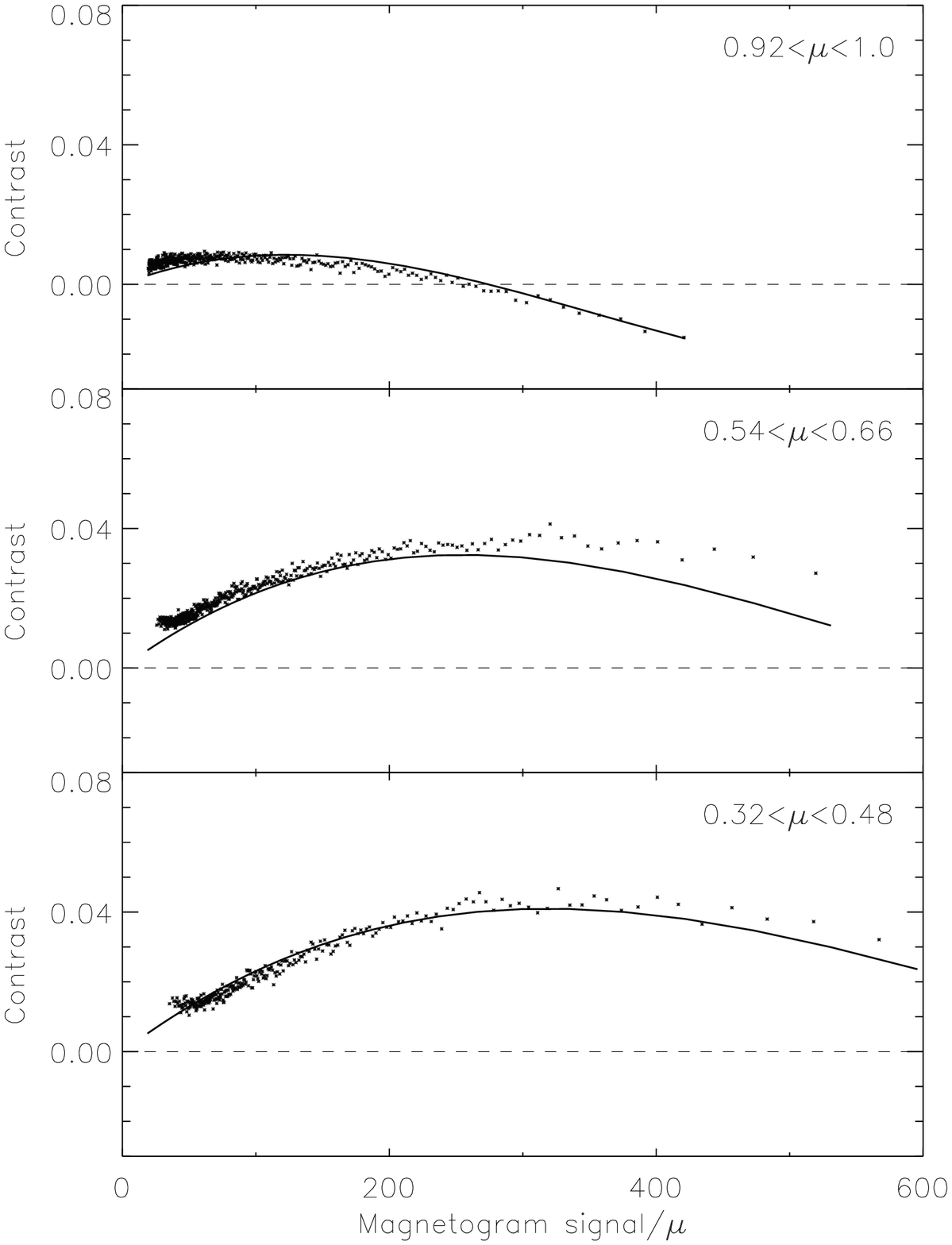}}
\vspace{0.4cm}
\caption{The same as Fig.~\ref{fig6}, but for cuts along the $B$/$\mu$-axis
(solid curves) made at three positions on the solar disk. Dots represent
measured contrasts. The plotted curves represent the same $\mu$ ranges as those
of the data points. Every dot represents 200 data points.}
\label{fig7}
\end{figure}

To better estimate how this analytical surface fits the behaviour of the
measured contrasts, we have sliced the surface in both directions, $\mu$ and
magnetograph signal, and then compared the result with the measured values. In
Fig.~\ref{fig6} the fitted surface is sliced along the $\mu$-axis (solid
curve), at three sample magnetic signal ranges, representative of low (top
panel), medium (middle panel) and high (lower panel) $B$/$\mu$ values. Dots
represent measured contrasts. To avoid very crowded plots each dot represents
250 (top panel), 100 (middle) and 25 (bottom) data points, respectively. The
different amounts of binning reflect the non-uniform distribution of points
over the $B$/$\mu$ range. The multivariate regression surface fits quite well
the plotted dependence of the contrast, although minor deviations are visible
at small $B$/$\mu$. Fig.~\ref{fig7} shows slices of the modeled surface along
the $B$/$\mu$ axis (solid curves) and the corresponding binned data (dots), at
three sample positions on the solar disk, from disk center (top panel) to near
the limb (lower panel). The fitted curves now deviate somewhat more from the
data points, most significantly for \mbox{$0.54<\mu<0.66$}, where the
discrepancy can reach 0.01 in contrast.

The multivariate analysis yields to an expression for the contrast of
photospheric bright features, $C_{\mathrm{fac}}(\mu, B/\mu)$, that cannot be
directly compared with previous studies because, to our knowledge, no similar
work has been done before. Quadratic functions have already been used by other
authors (e.g. Foukal \cite{f81}) to fit the dependence of the facular contrast
on position over the disk, although most of them use a function of the form
$C_{\mathrm{fac}}(\mu)=b(1/\mu -a)$ (Chapman \cite{chap80}). A quadratic
function agrees quite well with the CLV proposed by the hot wall model. A cubic
function has been used for fitting the dependence of the contrast on magnetic
strength. In this case we do not have a physical reason, only the goodness of
the fit with respect to other bivariate functional dependences tried (see
Figs.~\ref{fig6} and \ref{fig7}) and the requirement to force the contrast
through zero for a disappearing magnetic signal. We suspect that, in order to
obtain a better empirical description of the dependence of facular contrast on
$\mu$ and $B$/$\mu$, a larger number of free parameters is required.

The dependence of the peak of $C_{\mathrm{fac}}$ on $B/\mu$ is shown in
Fig.~\ref{fig8}. The $\mu$-values at which $C_{\mathrm{fac}}$ peaks,
$\mu_{\mathrm{max}}$, have been represented against the corresponding magnetic
signal in Fig.~\ref{fig8}a. Fig.~\ref{fig8}b shows the peak $C_{\mathrm{fac}}$
values reached, $C_{\mathrm{fac}}^{\mathrm{max}}$ (see Fig.~\ref{fig3}),
plotted as a function of $B/\mu$. Finally Fig.~\ref{fig8}c shows
$C_{\mathrm{fac}}^{\mathrm{max}}/(B/\mu)$ plotted against $B/\mu$ or, in other
words, the dependence on the magnetic signal of the specific contrast per unit
of magnetic flux. Errors in $\mu$ are estimated from the difference between the
peak of the best-fit curves in Fig.~\ref{fig3} and the peak obtained directly
from the data points. Error bars in $B/\mu$ correspond to the size of the
$B/\mu$-intervals used in Fig.~\ref{fig3}. 

A linear regression adequately describes the dependence of $\mu_{\mathrm{max}}$
on $B/\mu$ for the precision achievable with the current data. The best fit
straight line (solid line in Fig.~\ref{fig8}a) is 

\begin{equation} 
\mu_{max}=5.60\,10^{-1}-4.97\,10^{-4} (B/\mu).
\end{equation}

\noindent Thus, $\mu_{\mathrm{max}}=0.56\pm0.02$ when $B/\mu$ tends to zero,
and $\mu_{\mathrm{max}}=0$ for $B/\mu\approx1120$ \mbox{G} from extrapolations
of this curve. Fig.~\ref{fig8}c implies that the contrast per unit of magnetic
signal decreases strongly with increasing magnetogram signal. Since individual
flux tubes are not resolved by MDI, we cannot infer the intrinsic contrast of a
flux tube from Fig.~\ref{fig8}b, which obviously shows the same pattern as
Fig.~\ref{fig4}. However, by normalizing by $B/\mu$ we obtain a quantity that
is roughly proportional to the intrinsic brightness of the flux tubes (assuming
that the field strength of the elemental magnetic flux tubes lies in a narrow
range as mentioned in Sect.~\ref{dataproc}).


\section{Discussion}
\label{discu}
\subsection{Comparison with previous observations}

Comparison with other contrast observations is not easy because of the
differences in the selected wavelength, spatial resolution, range of studied
heliocentric angles, magnetic filling factor and size of the analyzed features.
All these factors contribute to the scatter between the existing contrast
measurements. 

Our results differ from earlier observations of the contrast of bright
features, specially when considering magnetic signals $B/\mu > 200$ \mbox{G} at
disk center. Previous measurements of disk center facular contrast have
frequently yielded positive values, although they usually were close to zero.
Thus, from multi-color photometric images, Lawrence (\cite{l88}) measures
$C_{\mathrm{fac}}\sim0.005$ at disk center, and Lawrence et al.
(\cite{lawchapher88}) find $C_{\mathrm{fac}}=0.007\pm0.001$. In fact, our
results agree better with those of Topka et al. (\cite{topka92},
\cite{topka97}) and Lawrence et al. (\cite{lawtopjo93}) despite the difference
in spatial resolution and studied wavelength. For
\mbox{$200\,\mbox{G}\leq(B/\mu)\leq600\,\mbox{G}$} the agreement is also
surprisingly good. Nevertheless, these authors distinguish between active
regions and quiet Sun. For active regions they always measure a negative
contrast around disk center (for $\mu=0.97$ and $\mu=0.99$) irrespective of the
magnetogram signal, while we get slighly positive contrast values for
$B/\mu\leq200$ \mbox{G} in agreement with their results for the network. Since
at these field strengths most of our signal originates in the network, this
agreement is probably not surprising.

\begin{figure} 
\vspace{-1.cm} 
\hspace{-0.2cm}
\resizebox{10cm}{13cm}{\includegraphics{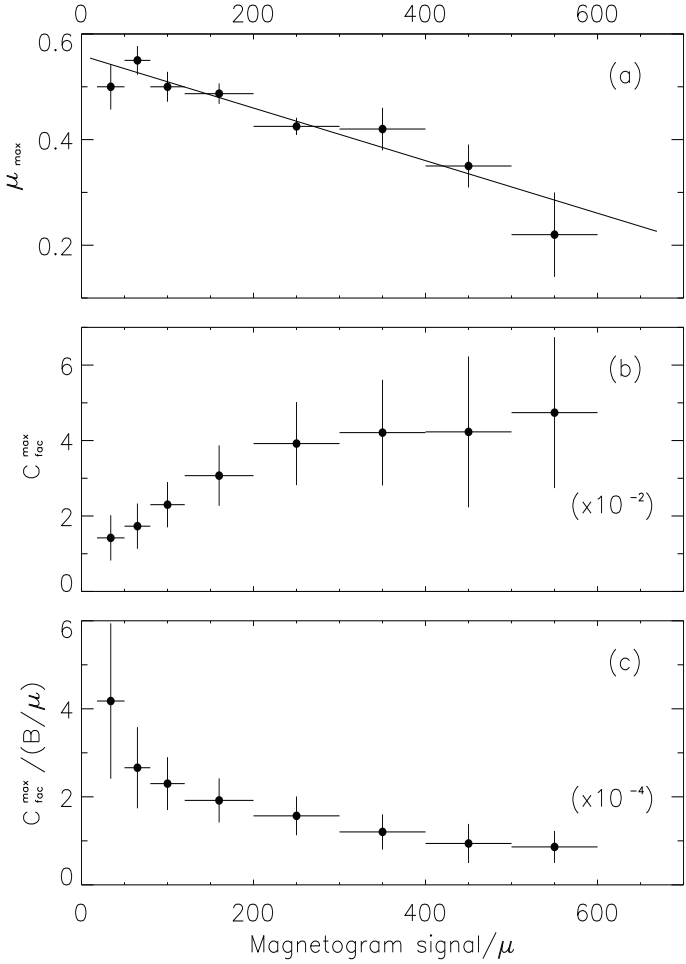}} 
\vspace{1cm} 
\caption{Dependence on the magnetic flux per pixel, $B/\mu$, of: {\bf a)}
$\mu_{\mathrm{max}}$; {\bf b)} $C_{\mathrm{fac}}^{\mathrm{max}}$ times
$10^{2}$; {\bf c)} $C_{\mathrm{fac}}^{\mathrm{max}}/(B/\mu)$ times $10^{4}$.
See text for details.} 
\label{fig8}
\end{figure}

Chapman \& Klabunde (\cite{ck82}) claim that the contrast shows a sharp
increase near the limb (and even fit a $\mu^{-1}$ dependence). We find that
$C_{\mathrm{fac}}$ peaks between $\mu=0.5$ and $\mu=0.2$, depending on the
magnetic strength of the signal, and then decreases towards the limb
(Fig.~\ref{fig3}). Libbrecht \& Kuhn (\cite{lk84}, \cite{lk85}) also find this
behaviour; however, they give $\mu\leq0.2$ for the peak of the contrast. Wang
\& Zirin (\cite{wz87}) and Spruit's hot wall model also give a similar value
for the $\mu$ at which $C_{\mathrm{fac}}$ peaks. It is worth noting that
Libbrecht \& Kuhn (\cite{lk84}, \cite{lk85}) and Wang \& Zirin (\cite{wz87}) do
not take into account the magnetic field of the observed feature, which makes
the comparison between our results and theirs more difficult, as the CLV
obtained is different when features are selected according to their brightness
rather than the magnetogram signal. In the former case there is a bias towards
brighter features. Our results indicate that the higher the magnetic signal,
the smaller the $\mu$-value at which the contrast peaks (see Fig.~\ref{fig8}a)
so that network-like features dominate at disk center and features with
increasingly large $B/\mu$ closer to the limb. This should move the peak of the
contrast to smaller $\mu$ when the brightest features are searched for, than
when magnetograms are used to identify faculae. Finally, it should be pointed
out that, for increasingly smaller $B/\mu$ values the contrast becomes
increasingly independent of $\mu$; this agrees with the conclusion of Ermolli
et al. (\cite{ermo99}) that the network contrast is almost independent of
$\mu$.

It is remarkable that an expression as given by Eq.~(\ref{eq2}) reproduces the
dependence of the contrast of bright features on their position ($\mu$) and on
the magnetic flux per pixel ($B$/$\mu$), within the range
\mbox{$0.1\leq\mu\leq1$} and \mbox{$17\,\mbox{G}\leq(B/\mu)\leq630\,\mbox{G}$}.
A relative accuracy of better than 10\% is achieved almost everywhere within
this domain. However, this multivariate analysis is only a first step and
considerable further work needs to be done, since two other relevant parameters
for the contrast, namely the wavelength and the spatial resolution, are kept at
fixed values (those prescribed by MDI) in our analysis. A 4-dimensional data
set is thus needed. A first step was taken by Lawrence et al.
(\cite{lawtopjo93}), who compared observations from different instruments. At
least some further progress in this direction can be achieved by employing MDI
high resolution data, although off-center pointing is required.

\subsection{Comparison with flux-tube models}

MHD models including self-consistent energy transfer predict that small flux
tubes (diameters smaller than 300 km) appear bright at disk center but with
decreasing contrast near the limb; somewhat larger tubes are predicted to
appear dark at disk center but bright near the limb, and finally, very large
flux tubes (pores and sunspots; not considered in this study) are predicted to
be dark everywhere (e.g., Kn\"olker \& Sch\"ussler \cite{ks88}, \cite{ks89}).
In such models the contrast at $\mu\approx0.1$ is largely determined by the
brightness of the bottom of the flux tube (and the brightness of its
surroundings, e.g. granular down flow lanes), while the CLV of the contrast is
strongly influenced by the visibility of the hot walls. The bottom of a flux
tube is defined as the horizontal optical depth unity surface in the interior
of a flux tube. 

Our results are qualitatively in accordance with this prediction if we make two
reasonable assumptions. First, the network and facular features are composed of
a mixture of spatially unresolved flux tubes of different sizes. Second, the
average size of the flux tubes increases with increasing magnetogram signal or
filling factor. Under these assumptions the upper panels of Fig.~\ref{fig3}
refer to, on average, small flux tubes which dominate the network, while the
lower panels of that figure refer to larger tubes mostly present in AR faculae.
In our study the contrast always has a minimum at $\mu=1$ and increases with
decreasing $\mu$ (as part of the hot wall becomes visible), until a maximum
when the contrast peaks (the maximum surface of the hot wall is seen). Closer
to the limb the contrast decreases as less wall surface is exposed. There are,
however, clear differences between small $B/\mu$ network flux tubes and tubes
found in AR faculae, i.e. regions with large $B/\mu$. Network tubes are bright
everywhere on the solar disk and exhibit a low contrast (Fig.~\ref{fig8}b), but
a high specific contrast (Fig.~\ref{fig8}c). This implies that network flux
tubes are brighter than AR flux tubes and partly reflects the fact that network
flux tubes are hotter than AR tubes (e.g., Solanki \& Brigljevi\'{c}
\cite{solbri92}; Solanki \cite{solanki93}). The greater brightness at large
$\mu$ implies that network flux tubes have a hotter bottom than larger flux
tubes. Since this is also true at $\mu\leq0.6$, it suggests that the walls of
smaller tubes, or of tubes in regions with lower filling factor, are hotter as
well. This is in agreement with the theoretical finding of Deinzer et al.
(\cite{deinz84b}) that the inflow of radiation into the tube leads to a cooling
of the surroundings and a lowering of the temperature of the walls. This
temperature reduction is indeed predicted to be greater for larger flux tubes
(Kn\"olker \& Sch\"ussler \cite{ks88}). 

A mixture of flux-tube sizes at a given $B/\mu$ is needed because the CLV of
$C_{\mathrm{fac}}$ at small $B/\mu$ does not agree with the predictions for any
size of flux tube. The model flux tubes are all bright over only a relatively
small range of $\mu$ values. Hence the mixture of flux tube sizes is needed in
order to produce a relatively $\mu$-independent contrast, as exhibited by
magnetic features at small $B/\mu$. As can be seen in Fig.~\ref{fig3}, the
contrast shows a more pronounced CLV as tube size increases, in accordance with
the hot wall model, and larger tubes have a negative contrast at disk center,
as predicted. The high specific contrast of small $B/\mu$ features
(Fig.~\ref{fig8}c), and the fact that their contrast is positive over the whole
solar disk indicates that a change in the magnetic flux of the network has a
much larger contribution to the change of the irradiance than a similar change
in flux in active regions.

From Fig.~\ref{fig8}a we can determine the heliocentric angles that make the
contrast peak, $\theta_{\mathrm{max}}$. For the intervals displayed on
Fig.~\ref{fig3}, $\theta_{\mathrm{max}}$ is $63\degr$, $55\degr$, $58\degr$,
$62\degr$, $66\degr$, $68\degr$, $72\degr$ and $77\degr$, respectively.
Assuming the hot wall model with a simplified cylindrical geometry for the flux
tubes, a Wilson depression $Z_{\mathrm{W}}$ of 150 km (Spruit \cite{spruit76})
and the derived $\theta_{\mathrm{max}}$ values, it is possible to roughly
estimate the average value of the tube diameter for each magnetic range. Taking
into account that the maximum depth of the wall $Z_{\mathrm{W}}$ is seen when
the angle between the local vertical to the tube and the line of sight is equal
to the heliocentric angle, then the diameter D should be
$D=Z_{\mathrm{W}}\tan(\theta_{\mathrm{max}})$. Applying this approximation to
our observations, we obtain diameters of 290, 210, 240, 281, 334, 365, 460, and
650 km respectively, for the mentioned $\theta_{\mathrm{max}}$ values and their
respective magnetic ranges. These diameters are estimated to be uncertain by
approximately a factor of two. For example, uncertainties in $Z_{\mathrm{W}}$
translate into proportionate relative uncertainties in D.

Finally, we wish to draw attention to the wiggle of the measured contrasts
around $\mu=0.95$ in Figs.~\ref{fig3} and \ref{fig6}. This can be observed at
all magnetic strengths. At $\mu=1$ the contrast has a minimum value, then
increases, descending a bit later for still smaller $\mu$'s before finally
increasing slowly towards the limb. Topka et al. (\cite{topka92}) show some of
these variations very close to disk center in Fig. 5 of their paper. They argue
that such variations are partly due to the inclination of the flux tubes of
opposite polarities toward each other in the active region they observe.
However, we average over many network elements on multiple days and the
persistence of such a structure is surprising, in particular also for small
$B/\mu$ values where the statistics are extremely good. 


\section{Conclusions}
\label{conclu}

The results presented in this work indicate that the CLV of the continuum
contrast of magnetic features changes gradually with magnetogram signal (or
magnetic filling factor), such that the contrasts of AR faculae and the network
exhibit a very different CLV, in general agreement with the results of Topka et
al. (\cite{topka92}, \cite{topka97}) and Lawrence et al. (\cite{lawtopjo93}). A
possible reason for the difference is that the populations of magnetic flux
tubes found in the two kinds of features are, on average, different in size, in
agreement with the conclusions of earlier investigations (Grossmann-Doerth et
al. \cite{grossdo94}; Keller \cite{keller92}). 

Stronger magnetogram signals, corresponding to wider flux tubes on average
(Grossmann-Doerth et al. \cite{grossdo94}), appear dark at disk centre, but
bright near the limb, while the weakest signals (on average na\-rro\-wer flux
tubes) are almost equally bright at disk centre and near the limb. This result
is in good agreement with the predictions of theoretical flux tube models
(Deinzer et al. \cite{deinz84a}, \cite{deinz84b}; Kn\"olker et al.
\cite{ketal88}; Kn\"olker \& Sch\"ussler \cite{ks88}) if there is a
distribution of flux tube sizes present within an MDI pixel. Because network
elements are bright over the whole solar disk their contribution to irradiance
variations is significant and needs to be taken into account when
reconstructing variations of the total solar irradiance.

One advantage of the present investigation relative to that of Topka et al.
(\cite{topka92}, \cite{topka97}) is that by using full disk MDI data we have a
result for a very well defined spatial resolution, so that any models derived
on the basis of these results can be directly used for reconstructing total and
spectral solar irradiance, as for instance measured by VIRGO (Fr\"ohlich et al.
\cite{froh95}), without further adjustment.

A new result of this work is that, with a simple expression, we can predict the
contrast of a bright feature, from network and small tubes to faculae of
different sizes, given its position and magnetic strength within a certain
range, and reproduce simultaneously the $C_{\mathrm{fac}}(\mu)$ and
$C_{\mathrm{fac}}(B/\mu)$ dependences.

In a next step the dependence of the contrast on wavelength (for given $\mu$
and $B/\mu$) must be determined, as well as the dependence on spatial
resolution. The later dependence is of particular interest also because the
investigations of Lawrence et al. (\cite{lawtopjo93}) and Topka et al.
(\cite{topka97}) give similar values of the contrast near $\mu=1$ as we find,
although the spatial resolution of the La Palma data employed by these authors
is almost an order of magnitude higher than that of the MDI full disk data
(0.5\arcsec versus 4\arcsec). Closer to the limb Topka et al. (\cite{topka97})
obtain contrasts a factor of two higher. Whether this is due to the different
wavelengths observed or has another source needs to be investigated.


\begin{acknowledgements}

AO acknowledges financial support from the DURSI (Ge\-ne\-ra\-li\-tat de
Catalunya) grant 2001\,TDOC\,00021, as well as partial financial support from
the Max-Planck-Institut f\"ur Aeronomie, from E. Ortiz and S. Carbonell, and
from the European Space Agency (under contract ESA-ESTEC 14098/99/NL/MM). VD
acknowledges partial financial support from the DURSI (Generalitat de
Catalunya).

\end{acknowledgements}


\end{document}